# PrivAgE: A Toolchain for Privacy-Preserving Distributed Aggregation on Edge-Devices

Johannes Liebenow[1] · Timothy Imort[1] · Yannick Fuchs[1] · Marcel Heisel[1] · Nadja Käding[2,3] · Jan Rupp[2,3] · Esfandiar Mohammadi[1]

**Abstract**
Valuable insights, such as frequently visited environments in the wake of the COVID-19 pandemic, can oftentimes only be gained by analyzing sensitive data spread across edge-devices like smartphones. To facilitate such an analysis, we present a toolchain called PrivAgE for a distributed, privacy-preserving aggregation of local data by taking the limited resources of edge-devices into account. The distributed aggregation is based on secure summation and simultaneously satisfies the notion of differential privacy. In this way, other parties can neither learn the sensitive data of single clients nor a single client's influence on the final result. We perform an evaluation of the power consumption, the running time and the bandwidth overhead on real as well as simulated devices and demonstrate the flexibility of our toolchain by presenting an extension of the summation of histograms to distributed clustering.

**Keywords** Distributed and Secure Aggregation · Edge-Devices · Differential Privacy · Acoustic Scene Classification (ASC) · Clustering

## 1 Introduction

Analyzing huge amounts of data can bring valuable insights in various scenarios. In most cases, such an amount of data is distributed among multiple clients, specifically edge-devices. Edge-devices are devices like smartphones or smartwatches that are limited in their resources, but provide a multitude of sensors, which can be used to collect all sorts of data. However, an increasing awareness of privacy concerns, e.g. supported by the general data protection regulation in the EU [6], requires that the privacy of individuals is protected. To still be able to maximize the value of sensory data while simultaneously providing the necessary degree of privacy protection, we see two approaches, namely privacy-preserving federated learning or specialized aggregation schemes. To the best of our knowledge, frameworks for privacy-preserving federated learning oftentimes do not provide any code [10, 11], or the provided code can only be used to reproduce experimental results and set up a local implementation [3, 7]. The same holds for privacy-preserving aggregation schemes specialized on specific use-cases like heavy hitters [12] or distributed clustering [4]. Even if an implementation is provided, it can only be used to create a local setup.

✉ Johannes Liebenow
j.liebenow@uni-luebeck.de

Timothy Imort
timothy.imort@student.uni-luebeck.de

Yannick Fuchs
yannick.fuchs@student.uni-luebeck.de

Marcel Heisel
marcel.heisel@student.uni-luebeck.de

Nadja Käding
nadja.kaeding@uni-luebeck.de

Jan Rupp
jan.rupp@uksh.de

Esfandiar Mohammadi
esfandiar.mohammadi@uni-luebeck.de

[1] Institute for IT-Security, University of Lübeck, Lübeck, Germany

[2] University Hospital Schleswig-Holstein, Campus Lübeck, Germany

[3] Department of Infectious Diseases and Microbiology, University of Lübeck, Lübeck, Germany

Although frameworks for federated learning and specialized aggregation schemes are an important building block, we argue that there is also the need for a toolchain which can be used to implement the basic setup of distributed learning. This includes an app which is compatible to edge-devices and also a server which coordinates the learning process. Such a setup should lay the foundation for providing the necessary degree of privacy protection and for taking into account the limited resources of edge-devices.

In this light, we propose a toolchain called PrivAgE that provides the basic setup for distributed learning on edge-devices and enables its rapid realization. This setup includes an app, a server and a communication protocol. First, a user installs the app on their edge-device and the app starts a data collection phase. At a certain point, the data collection stops and the collected data gets pre-processed into a summable format. Then, the server coordinates a secure aggregation to obtain a global result from all the local, pre-processed data of clients. After the aggregation, the server publishes the global result on a website to facilitate further analyses.

To enable secure aggregation, we provide an implementation of a state-of-the-art secure summation protocol [2] such that the server and other parties cannot learn individual inputs. Clients introduce random noise to their local data before the aggregation. By making use of the popular privacy notion *differential privacy* [5], the local noise addition in combination with the security guarantees of the secure summation protocol suffice to hide the influence of clients in the aggregated sum, i.e. to protect the privacy of clients. The implementation of the data collection, the exact format of the aggregated data and safeguarding differential privacy is tailored to the specific use-case.

This work is a proof of concept of our toolchain for a given use-case. The use-case is about collecting information about frequently visited environments in a pandemic like COVID-19 [1]. Using our toolchain, clients locally record audio files, determine the surrounding environment via machine learning called *acoustic scene classification* and aggregate the resulting environment labels to a local histogram. To further enrich the local data, clients also include the number of Bluetooth devices in their surrounding into the local histogram. Then, random noise is added to their local statistic and a secure aggregation takes place based on the provided secure summation protocol. Afterwards, the server publishes the final histogram to enable a further analysis by experts. The summed up noise in combination with the guarantees of the secure summation suffice to protect the privacy of single clients' data.

Finally, we evaluate the power consumption, the running time and the bandwidth overhead of the most consumptive parts of our toolchain applied to the use-case of environment labels and thereby demonstrate its compatibility with edge-devices. Additionally, we discuss and demonstrate an extension of our toolchain to distributed, differentially private clustering. This naturally extends our use-case in the scenario of the COVID-19 pandemic for extracting not only the most frequent environment labels but sequences of labels. We present how we generated synthetic data, demonstrate that such sequences can indeed be clustered and discuss how this could be implemented.

*Contribution*

– We present PrivAgE a toolchain for distributed and secure aggregation of sensory data explicitly designed for edge devices.
– We use a state-of-the-art secure summation protocol to prevent any external party and even the central server to learn individual inputs and apply the notion of differential privacy to hide the presence of single clients' data in the final result.
– We demonstrate the flexibility of our toolchain by directly incorporating a use-case for improving the information flow in scenarios like a pandemic and discuss a potential extension to distributed clustering. Our source code is available at https://github.com/UzL-PrivSec/PrivAgE.

## 2 Parties

We consider the following parties in the context of our toolchain:

*Users* A user collects and stores sensitive data on their device. Thus, users have a vested interest in protecting their own data from unauthorized individuals. Additionally, a device is limited in its resources due to natural constraints imposed by, e.g., smartphones.

*Server* The server acts as an intermediary in our toolchain by providing necessary hyper-parameters and coordinating the aggregation. In contrast to clients, the server is equipped with high computational resources to efficiently perform the aggregation process.

*Third Parties* Other parties, such as public health officials, researchers, or users, are permitted to analyze the published results. However, they should not be able to learn single clients' data (security) or whether a specific client was part of the aggregation (privacy).

## 3 Edge-Device Toolchain

In this section, we give an overview over the individual steps of our aggregation toolchain and present design choices as well as details on the implementation. The specific realization of the individual steps depends on our use-case of extracting frequently visited environments in a pandemic.

## 3.1 Overview

*Local Data Collection & Pre-Processing* The app of our toolchain automatically collects environment labels over a specific period of time by using the microphone of the device and a machine learning model trained on acoustic scene classification. The environment labels are represented in the form of a histogram. To preserve privacy when dealing with histograms, clients inject random noise into each count of the histograms [5].

*Secure Aggregation* The locally noised histograms are securely aggregated using a specific secure multiparty computation (SMPC) protocol. This protocol allows clients to collaboratively compute an aggregated sum without revealing individual inputs to the server. After aggregating all local histograms, the noise introduced by each user separately suffices to hide the influence of a single user on the aggregated result. Thus, third parties remain oblivious to individual contributions, safeguarding the privacy of the users.

*Publishing* After the server has aggregated the local histograms in a secure and privacy-preserving way, the resulting global histogram is ready to be published. Besides coordinating the secure aggregation, the server also hosts a website where the final results are made public. Interested third parties such as public health officials or researchers can now gain insights into the overall patterns of visits to different environments.

## 3.2 Technical Building Blocks

This section presents details about the building blocks of PrivAgE, namely acoustic scene classification (ASC) for collecting environment labels, a state-of-the-art secure summation protocol and differential privacy to guarantee the necessary degree of privacy.

*Acoustic Scene Classification* The data collection step mainly consists of a machine learning task called acoustic scene classification. Our app regularly records audio files on a user's device. The files are used as input for a neural network trained to predict the environment in which the recording took place. We call this prediction the environment label. To enhance this information, we incorporate the number of Bluetooth devices present at the time of recording, which can be obtained using Bluetooth LE technology. By aggregating local histograms over environment labels and Bluetooth device counts, we generate statistics that are highly valuable to experts. Especially in scenarios like the COVID-19 pandemic, these statistics can contribute to a better understanding of infection waves.

*Secure Summation* We aim to ensure security for individual inputs in the aggregation. Therefore, we employ a secure summation protocol to aggregate the local histograms of individual users into a single global histogram. For this purpose, we utilize the protocol introduced in [2] which can handle scenarios where some clients drop out during the execution, which is a realistic consideration when dealing with edge-devices. We have implemented the standard version of the protocol in which the server acts honest-but-curious which means that the server follows the protocol but tries to infer as much information as possible. The protocol can also be extended in a way such that it is able to tolerate a malicious server cooperating with other malicious clients.

*Privacy Protection* To protect the privacy of clients who take part in the aggregation, we require the aggregation to satisfy the notion of differential privacy (DP).

**Definition 1** Differential Privacy (DP). A randomized algorithm $M : \mathcal{D} \to A$ is $\varepsilon$-differentially private if for any pair of databases $D_0, D_1 \in \mathcal{D}$ that only differ in a single element and all tests $S \subseteq A$ on the result of $M$ the following holds:

$\Pr[M(D_0) \in S] \leq \exp(\varepsilon) \cdot \Pr[M(D_1) \in S]$.

We consider a scenario in which a data set is distributed among clients. The data set contains data points which attributes can be correlated. Thus, when protecting the privacy of individuals, we aim to hide the influence of entire data points and not single attributes. We consider two neighboring versions of the distributed data set, one data set $D_0$ with all the data points and one data set $D_1$ where the data points of a single client are missing. One of the data sets is used as input for our aggregation algorithm $M$. An attacker then receives the output of $M$ and has to decide which data set was used as input. Formally (see Definition 1), we say that if the output distributions of the aggregation $M(D_0), M(D_1)$ regarding neighboring data sets are similar s.t. their ratio is bounded by $\exp(\varepsilon)$, then the algorithm satisfies DP. In this way, the degree of privacy protection can be controlled via the parameter $\varepsilon$ called privacy budget. Generally, it can be assumed that with $\varepsilon \leq 1$ a reasonable privacy protection is ensured. To satisfy DP, an algorithm has to be randomized and for further details we refer to [5]. Intuitively, this means that even if an attacker has knowledge over the entire data set except for the data points of a single user, the influence of this user stays hidden and thus the privacy of this user can be protected. To hide the influence of a single users' data, their worst-case influence on the final result has to be bounded from above by a finite value. We call this value sensitivity. When working with environment labels, we restrict the local count per environment label and the count of Bluetooth devices to a maximum of $c$ and $b$, respectively. Now, the influence of a single client is bounded and by adding Laplacian noise scaled by $\frac{c}{\varepsilon}$ and $\frac{b}{\varepsilon}$, the summation of environmental labels and number of Bluetooth devices preserves $\varepsilon$-DP [8].

In our distributed scenario, the server first publishes the required privacy budget $\varepsilon$ and then performs an aggregation of all clients' individual data. Although clients locally add the required amount of noise to their counts, directly aggregating all noised counts does not preserves privacy since other parties and the server have direct access to the individual noised inputs. Therefore, we use a secure summation protocol such that the server only obtains the sum of all individual inputs. In this way, we effectively simulate a trusted aggregator which enables the aggregation to satisfy DP.

To account for potential dropouts, where a proportion $\theta$ of users may drop out, users are required to scale their noise by $\frac{1}{1-\theta}$. This precaution ensures that even in the worst-case scenario the aggregated noise remains sufficient to achieve differential privacy.

## 4 Evaluation

### 4.1 Implementation Details

The implementation of PrivAgE involves several components. The app was developed using Android Studio in Java and Kotlin, while the SMPC functionality was implemented using the Java Native Interface in C++. On the server side, we implemented the secure summation part in C++. Both the app and server components of the protocol rely on the libraries "boost" and "cryptolib" to facilitate their functionality. The web services and the underlying database aspects are implemented using Python3 and the framework Django.

### 4.2 Experiments

To demonstrate the efficiency of our aggregation toolchain, we conduct experiments of specific parts of the toolchain. The results show that being part of the aggregation only introduces a moderate overhead to a client's device. We perform all measurements on Google Pixel 5 smartphones and to perform secure aggregation, we use simulated clients.

*Running Time* As secure aggregation constitutes the most to the running time, we measure the running time of our implementation of the protocol. We let the secure summation run 50 times for three different numbers of simulated clients. The results can be seen in Table 1. Our implementation is practical for a large amount of users: The running time for secret sharing and the pseudo-random generator (PRG) evaluation roughly doubles for every order of magnitude. This means our implementation succeeds in maintaining a sub-linear running time (in the number of total clients) which is the main reason for the efficiency of the protocol.

*Power Consumption* Acoustic scene classification (ASC) and secure summation are the central components in our use-case which is why we measure the respective power consumption. We first measure the baseline power consumption on a Google Pixel 5 smartphone without any active apps. To simulate the application of collecting environment labels in the every day life, we activate the ASC every 5 minutes and once a minute to underline the high consumption of performing inference with a neural network. We also include power measurements for plain secure aggregation of 100 users which consists of invoking of the secure summation protocol once per hour.

The plot in Fig. 1 displays the power consumption of the four scenarios. It shows that performing ASC once a minute drastically increases the power consumption. However, when comparing the daily routine to the baseline, the power consumption only increases by 5%. This is further reduced if ASC is omitted and the secure summation protocol gets invoked once per hour. In summary, the results show that secure aggregation and ASC without too many activations only increase the power consumption by a tolerable amount.

*Bandwidth* To measure the amount of data a smartphone receives and sends when being part of our toolchain, we make use of a profiler, an in-built tool of Android Studio, for measuring different system resources. A client first collects

**Table 1** Running time on simulated clients (in seconds) for Shamir secret sharing, PRG expansion (AES in counter mode) and the entire protocol. We set the hyper-parameters $\gamma = 1/20, \delta = 1/3, \sigma = 40, \eta = 30$ accordingly. For further details, we refer to [2]

| Users | Neighbours | Sharing | PRG Eval | Total |
|---|---|---|---|---|
| $10^3$ | 83 | 0.017 | 0.033 | 24.5 |
| $10^4$ | 103 | 0.033 | 0.078 | 27.43 |
| $10^5$ | 109 | 0.061 | 0.112 | 28.41 |

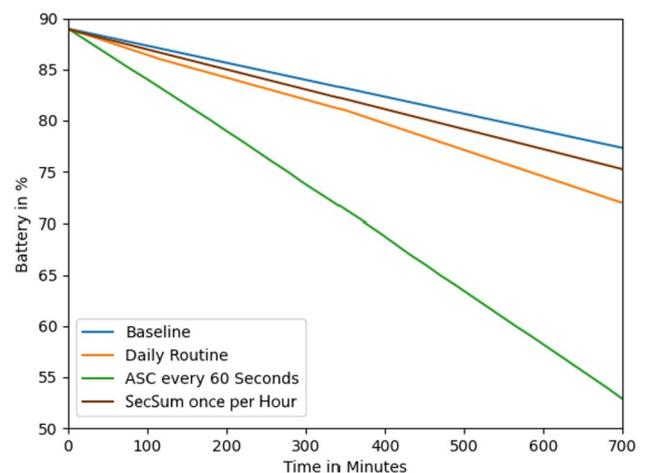

**Fig. 1** Effects of PrivAgE on the battery level on a Google Pixel 5 over 12 hours in idle-mode (*blue*), acoustic scene classification (ASC) every 5 minutes (*orange*), ASC every minute (*green*) and plain secure summation (SecSum) executed once per hour (*brown*). While ASC every minute has a large impact on the battery level, executing ASC and SecSum not every minute only introduces a small overhead

a few environment labels, creates a histogram and takes part in a secure aggregation of local histograms with around $10^4$ participants in total which requires a single invocation of the secure summation protocol. In the process, the client is a real smartphone and the other participants are simulated clients.

The results show that a client receives and sends around 3 MB of data. This means our implementation requires around 6 MB of data. For comparison: To browse on Instagram for 5 minutes requires on average 35 MB of data which is more than 5× of our traffic. The experiments on the bandwidth show that the collection of environment labels in combination with secure aggregation only lead to a small amount of additional traffic.

## 5 Clustering Short Traces

To demonstrate that our toolchain is not restricted to the aggregation of histograms, we propose an interesting extension to distributed clustering as future work. This is also an extension of our use-case, because with clustering, we are interested not only in frequently visited environments but in frequently occurring sequences of environments we call traces. We first demonstrate how to generate a synthetic data set of traces and present results on how differentially private clustering performs on these traces. Finally, we discuss the implementation of the distributed analysis of traces.

### 5.1 Synthetic Trace Generation

Since we are not aware of any real-world data set containing traces, we generate synthetic data first. Specifically, we use random walks to generate synthetic traces. The resulting data set contains two types of traces: The first type represents repetitive daily routines, characterized by common patterns that occur frequently. These patterns have a high occurrence rate and mirror societal behavior. The second type of traces captures individual movement patterns that occur from time to time but are not representative for the population and are regarded as noise.

### 5.2 Evaluation

To demonstrate that traces can be aggregated by using a differentially private clustering algorithm, we apply such an algorithm in the central scenario. Specifically, we use the diffprivlib [9] Python library, which provides a differentially private clustering algorithm based on KMeans. Our objective is to identify traces that represent frequently occurring daily routines. In our evaluation, we vary the privacy budget $\varepsilon$ and the number of random walks, both affecting the accuracy of the clustering. Each random walk can be considered as a single individual generating environment labels through their daily activities, resulting in more traces. The results of our evaluation are depicted in Fig. 2. For benchmarking purposes, we also compare the differentially private KMeans clustering implementation to a non-differentially private version.

With an increased number of random walks (samples), the clustering accuracy gets better. For a very strong privacy protection ($\varepsilon = 0.1$) the accuracy is low. When increasing $\varepsilon$ to 1, the performance drastically improves to about 30%. Further increasing $\varepsilon$ only improves the accuracy marginally to about 43%. The non-private clustering has an accuracy of up to 78%. This demonstrates that frequently occurring traces can be obtained via clustering while simultaneously enforcing privacy protection.

### 5.3 Integration

To integrate the analysis of traces into our toolchain, one can utilize the differentially private distributed clustering algorithm LSH-Splits [4]. It can be implemented in a distributed manner solely based on secure summation which makes it compatible with our toolchain. The algorithm consists of two parts. The first part happens locally on each client and the second part at the server. Locally, clients project their data into a lower dimensional space by using a public projection which consists of a single matrix multiplication. Next, each client has to assign each of their data points to the closest point from a set of predefined reference points of size $k$. This involves $\mathcal{O}(k)$ distance calculations. Then, clients derive two special data structures from the closest reference points represented as a histogram and a vector and send both to the server. On the server side, the sum of vectors and the sum of histograms are used to compute a clustering for the distributed data set. Since this algorithm requires only a single round of communication, an integration into our toolchain would require only two invocations of the secure summation protocol, one for each of the local data structures. In summary, the LSH-Splits clustering algorithm can be integrated into our toolchain due to the negligible number of computations for clients and the compatibility of the aggregation operations with secure summation.

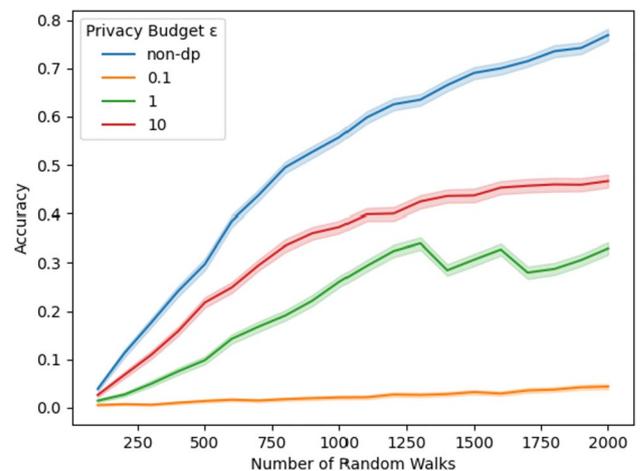

**Fig. 2** Clustering accuracy of differentially private clustering for identifying frequently occurring traces. We use different privacy budgets and compare to a non-private K-Means baseline. With an increased number of random walks (more samples) the accuracy gets better and for a privacy budget of 1 we can still achieve a reasonable accuracy

In this scenario, clients create traces from local environment labels obtained via acoustic scene classification and join the distributed clustering when initiated by the server with their traces as data points. Due to secure summation, the server cannot learn individual inputs and only obtains the final cluster centers. The information which can be gained from these centers is that each center directly corresponds to a frequent trace. In this way, local traces can be aggregated in a secure and privacy-preserving way with only a few adjustments to the underlying aggregation algorithm.

# 6 Conclusion

We introduce a toolchain called PrivAgE to enable the aggregation of sensitive data distributed across edge-devices. Our toolchain takes the limited resources of edge-devices into account and covers the process from collecting local data to the publishing of results. It is separated into two parts, an app and a server to coordinate the aggregation and publish the result on a website. The aggregation is based on a secure summation protocol and differential privacy. Thus, other parties cannot learn individual inputs nor their influence on the result.

Throughout this work, we present the toolchain in the light of a special use-case based on a machine learning model for acoustic scene classification, specifically in the context of the COVID-19 pandemic, to facilitate the analysis of frequently visited environments. To demonstrate the efficiency of our toolchain in this use-case, we measure the resource consumption on real and simulated devices in terms of power, time and bandwidth. In summary, collecting data locally and being part of the aggregation only introduces a moderate overhead to a client's device.

To further demonstrate the flexibility of our toolchain we present an extension from environment labels to sequences of labels and show that it can directly be extended to differentially private clustering to obtain frequently occurring sequences of environmental labels.

**Acknowledgements** We thank the anonymous reviewers for their helpful feedback. This work has been supported by the Volkswagen Foundation (PRIVEE), the DFG (CoContexts) and the BMBF (AnoMed and MLens).

**Funding** Open Access funding enabled and organized by Projekt DEAL.

**Data availability** There is no data coming with this publication.